\documentclass[
reprint,           
superscriptaddress,
amsmath,           
amssymb,           
aps,               
prd,               
notitlepage,       
longbibliography,  
floatfix,          
showkeys,          
]{revtex4-1}

\usepackage{tensor}     
\usepackage{graphicx}   
\usepackage[
colorlinks=true,        
citecolor=blue,         
linkcolor=blue,         
urlcolor=blue           
]{hyperref}             
\usepackage{bm}         
\usepackage{xcolor}     
\usepackage{tabulary}
\usepackage{color}      
\usepackage[section]{placeins} 

\renewcommand{\vec}[1]{\boldsymbol{#1}} 

\newcommand{\nc}{\newcommand*} 

\nc{\al}{\alpha}
\nc{\s}{\sigma}
\nc{\dt}{\delta}
\nc{\Dt}{\Delta}
\nc{\Ld}{\Lambda}
\nc{\p}{\partial}
\nc{\om}{\omega}
\nc{\Om}{\Omega}
\nc{\rd}{\mathrm{d}}
\nc{\Od}[1]{\mathcal{O}(#1)} 
\nc{\kp}{\kappa}

\def\({\left(}
\def\){\right)}
\def\[{\left[}
\def\]{\right]}
\def\e{\begin{equation}}
\def\q{\end{equation}}
\def\m{\begin{eqnarray}}
\def\n{\end{eqnarray}}

\nc{\Eq}[1]{Eq.~\eqref{#1}}     
\nc{\Fig}[1]{Fig.~\ref{#1}}     
\nc{\Table}[1]{Table~\ref{#1}}  
\nc{\Sec}[1]{Sec.~\ref{#1}}     

\nc{\Msun}{M_\odot}             
\nc{\fpbh}{f_{\mathrm{pbh}}}    
\nc{\fpbhn}{f_{\mathrm{pbh0}}}    
\nc{\mR}{\mathcal{R}} 
\nc{\seq}{\sigma_{\mathrm{eq}}}
\nc{\ogw}{\Omega_{\mathrm{GW}}}
\nc{\gpcyr}{\mathrm{Gpc}^{-3}\,\mathrm{yr}^{-1}}
\nc{\lvc}{LIGO/Virgo} 
\nc{\SNR}{\mathrm{SNR}} 
\nc{\mmin}{{m_{\mathrm{min}}}}
\nc{\mmax}{{m_{\mathrm{max}}}}
\nc{\Mmin}{{M_{\mathrm{min}}}}
\nc{\fmin}{{f_{\mathrm{min}}}}
\nc{\VT}{\mathrm{VT}}
\nc{\rhoGW}{\rho_{\mathrm{GW}}}
\nc{\vth}{\vec{\theta}}
\nc{\vd}{\vec{d}}
\nc{\vla}{\vec{\lambda}}
\nc{\Nobs}{N_{\mathrm{obs}}}
\nc{\av}[1]{\langle #1 \rangle} 
\nc{\km}{\mathrm{km}}
\nc{\Mpc}{\mathrm{Mpc}}
\nc{\Tobs}{T_{\mathrm{obs}}}
\nc{\Ntemp}{N_{\mathrm{temp}}}

\nc{\addref}{[\textcolor{red}{add ref}] } 
\nc{\eg}{\textit{e.g.~}}
\nc{\app}{\approx}
\nc{\hf}{\frac{1}{2}}
\nc{\discuss}{\textcolor{red}{Add discussion here!}}
\nc{\red}[1]{\textcolor{red}{#1}}

\nc{\mH}{\mathcal{H}}
\nc{\cs}{c_s^2}
\nc{\Sij}[1]{S_{ij}^{(#1)}}
\nc{\vi}[1]{v_i^{(#1)}}
\nc{\no}{\nonumber}
\def\<{\left\langle}
\def\>{\right\rangle}

\nc{\bk}{\bm{k}}
\nc{\bq}{\bm{q}}
\nc{\bp}{\bm{p}}
\nc{\bl}{\bm{l}}
\nc{\bx}{\bm{x}}
\nc{\be}{\mathbf{e}}
\nc{\mS}{\mathcal{S}}
\nc{\te}{\tilde{\eta}}
\nc{\tp}{\tilde{p}}
\nc{\tk}{\tilde{k}}
\nc{\tx}{\tilde{x}}
\nc{\tF}{\tilde{F}}
\nc{\tA}{\tilde{A}}
\nc{\mkpq}{|\bk-\bp-\bq|}
\nc{\mpq}{|\bp-\bq|}
\nc{\mkp}{|\bk-\bp|}
\nc{\mSi}[1]{\mS^{(#1)}({\bk, \eta})}
\nc{\vk}{\vec{k}}
\nc{\kstar}{k_*}
\nc{\xstar}{x_*}
\nc{\mpbh}{m_{\rm{pbh}}}
\nc{\cR}{\mathcal{R}}

\begin{document}
	
\title{The merger history of primordial-black-hole binaries}
	
\author{You Wu}
\email{youwuphy@gmail.com. Project 11847107 supported 
	by National Natural 
	Science Foundation of China. }
\affiliation{School of Mathematics and Physicsal sciences,
	Hunan University of Arts and Science, Changde, 415000, China}

	

\begin{abstract}
As a candidate of dark matter, primordial black holes (PBHs) have attracted 
more and more attentions as they could be possible progenitors of 
the heavy binary black holes (BBHs) observed by \lvc.
Accurately estimating the merger rate of PBH binaries will be crucial to
reconstruct the mass distribution of PBHs.
It was pointed out the merger history of PBHs may shift the merger rate distribution depending on the mass function of PBHs.
In this paper, we use $10$ BBH events from \lvc\ O1 and O2 observing runs to
constrain the merger rate distribution of PBHs by accounting the effect of
merger history.
It is found that the second merger process makes subdominant contribution to the 
total merger rate, and hence the merger history effect can be safely neglected.

\end{abstract}

\pacs{???}
	
\maketitle

\section{Introduction}
The direct detection of gravitational wave (GW) from a binary black hole (BBH)
coalescence \cite{Abbott:2016blz} has opened a new window of astronomy.
Over the past few years, ten BBH mergers have been reported by \lvc\ during 
the O1 and O2 observing runs \cite{Abbott:2016blz,Abbott:2016nmj,%
TheLIGOScientific:2016pea,Abbott:2017vtc,%
Abbott:2017gyy,Abbott:2017oio,LIGOScientific:2018mvr}.
The progenitors of these BBHs are still unknown and under intensively investigation (see \eg \cite{Bird:2016dcv,Sasaki:2016jop,Chen:2018czv,%
Fishbach:2017dwv,Clesse:2016vqa,Antonini:2016gqe,%
Inayoshi:2017mrs,Ali-Haimoud:2017rtz,Perna:2019axr,Kavanagh:2018ggo,%
Rodriguez:2015oxa,Rodriguez:2016kxx,Park:2017zgj,%
Belczynski:2014iua,Belczynski:2016obo,Woosley:2016nnw,%
Rodriguez:2018rmd,Choksi:2018jnq,2010AIPC.1314..291D,deMink:2016vkw}).
These \lvc\ BBHs present a much heavier mass distribution (in particular the 
source-frame primary mass of GW170729 event can be as heavy as
$50.2^{+16.2}_{-10.2} \Msun$ \cite{LIGOScientific:2018mvr}) than that 
inferred from X-ray observations
\cite{Wiktorowicz:2013dua,Casares:2013tpa,Corral-Santana:2013uua,%
Corral-Santana:2015fud},
which would challenge the formation and evolution mechanisms of astrophysical
black holes. 
One possible explanation for \lvc\ BBHs is the primordial black holes (PBHs)
\cite{Bird:2016dcv,Sasaki:2016jop,Chen:2018czv}
formed through the gravitational collapse of the primordial density
fluctuations \cite{Hawking:1971ei,Carr:1974nx}, 
which may accompany the induced GWs
\cite{Yuan:2019udt,Yuan:2019wwo,Chen:2019xse,Yuan:2019fwv}.
On the other hand, PBHs can also be a candidate of cold dark matter (CDM),
and the abundance of PBHs in CDM has been constrained by various experiments
\cite{Carr:2009jm,Barnacka:2012bm,Graham:2015apa,Niikura:2017zjd,%
Griest:2013esa,Niikura:2019kqi,Tisserand:2006zx,Brandt:2016aco,Gaggero:2016dpq,%
Ali-Haimoud:2016mbv,Blum:2016cjs,Horowitz:2016lib,Chen:2016pud,%
Wang:2016ana,Abbott:2018oah,Magee:2018opb,Wang:2019kaf,Chen:2018rzo,Chen:2019irf,%
Yuan:2019udt,Chen:2019xse}.

In order to account for the \lvc\ BBHs, the merger rate of PBH binaries has
been estimated to be $17 \sim 288~\gpcyr$ \cite{Chen:2018rzo}.
One should notice that theoretically there exist some uncertainties in
estimating the merger rate distribution of PBH binaries, and the estimation has
been continuing improved.
The merger rate of PBH binaries with monochromatic mass function has already
been given in \cite{Nakamura:1997sm,Sasaki:2016jop} for the case where two 
neighboring PBHs having sufficiently small separation can form a binary in
the early Universe due to the torque from the third nearest PBH.
These binaries would then evolve and coalesce within the age of the Universe 
and finally explain the merger events observed by \lvc\
\cite{Sasaki:2016jop}.
Later, the merger rate estimation is improved in \cite{Ali-Haimoud:2017rtz} 
by taking into account the torques exerted by all CDM (including all the PBHs
and linear density perturbations), but it is also assumed that all PBHs have
the same mass.
It is also pointed out in \cite{Ali-Haimoud:2017rtz} that the effects such as
encountering with other PBHs, tidal field from the smooth halo, 
the the baryon accretion are subdominant and can be neglected when estimating
the merger rate.

Various attempts have been made to estimate the merger rate distribution
of PBH binaries when PBHs have an extended mass function 
\cite{Chen:2018czv, Chen:2018rzo, Raidal:2018bbj, Raidal:2017mfl}.
In particular, a formalism to estimate the effect of merger history of PBHs
on merger rate distribution has been developed in \cite{Liu:2019rnx},
and it is argued that the multiple-merger effect may not be ignored if
PBHs have a power-law or a log-normal mass function by choosing some specific
parameter values of the mass function.
An accurate estimation of the merger rate will be crucial to infer the event
rate of \lvc\ BBHs and constrain the abundance of PBHs in CDM either through
null detection of sub-solar mass PBH binaries or the null detection
of stochastic GW background (SGWB) from PBH binaries.

In this paper, we will use the public available GW data from \lvc\ O1 and O2 
observations to estimate the merger rate distribution of PBH binaries with a
general mass function assuming all of \lvc\ BBHs are or primordial-origin. 
We find that the merger history effect makes no significant contribution to
the merger rate of PBHs and can be safely ignored.
The rest of this paper is organized as follows.
In \Sec{merger}, we review the calculation of merger rate distribution 
accounting for the merger history effect.
In \Sec{method}, we elaborate the data analysis method used to infer the
PBH populations from \lvc\ data.
In \Sec{result}, we present the results for PBHs with a power-law and 
a log-normal mass function respectively.
Finally, we summarize and discuss our results in \Sec{conclusion}.

\section{\label{merger}Merger rate distribution of PBHs}
In this section, we will briefly review the calculation of merger rate density
by closely following \cite{Liu:2019rnx}.
See also \cite{Chen:2018rzo}.
We denote the probability distribution function (PDF) of PBH masses by $P(m)$
which satisfies the following normalization condition
\e\label{norm}
    \int_{0}^{\infty} P(m)\, \rd m = 1.
\q 
Consequently the abundance of PBHs in the mass interval $(m, m+\rd m)$ is given 
by \cite{Chen:2018rzo}
\e 
    0.85 \fpbh\, P(m)\, \rd m,
\q 
where $\fpbh$ is the fraction of PBHs in CDM, and the coefficient $0.85$ accounts
for the fraction of CDM in non-relativistic matter.
Similar to \cite{Liu:2019rnx}, one may define a quantity $\mpbh$ as
\e\label{mpbh}
    \frac{1}{\mpbh} = \int \frac{P(m)}{m} \rd m.
\q 
Furthermore, the present average number density of PBHs with mass $m$ in the 
present total average number density of PBHs, $F(m)$, can be obtained by
\cite{Liu:2019rnx}
\e\label{Fm} 
    F(m) = P(m) \frac{\mpbh}{m}.
\q 
After some cumbersome derivations as presented in \cite{Liu:2019rnx}, 
one can get the total merger rate density, $\cR(t, m_i, m_j)$, 
of PBHs at cosmic time $t$ with masses $m_i \Msun$ and $m_j\Msun$ to be 
\e\label{cR}
    \cR(t, m_i, m_j) = \sum_{n=1} \cR_n(t, m_i, m_j),
\q 
where $\cR_n(t, m_i, m_j)$ is the merger rate density in the $n$-th merger process.
Then the total merger rate can be obtained by
\e\label{Rt}
    R(t) = \int \cR(t, m_i, m_j) \rd m_i \rd m_j = \sum_{n=1} R_n(t),
\q 
where
\e 
 R_n(t) = \int \cR_n(t, m_i, m_j) \rd m_i \rd m_j.
\q 
As demonstrated in \cite{Liu:2019rnx}, $\cR_{n+1}(t, m_i, m_j)$ is not 
necessarily be smaller than $\cR_n(t, m_i, m_j)$ (see Fig.~7 and Fig.~8 in 
\cite{Liu:2019rnx}).
However, $R_{n+1}(t)$ should be smaller than $R_n(t)$ as expected
\cite{Liu:2019rnx}.
Here, we only consider the merger history up to second-merger process.
The merger rate density of first-merger process, $\cR_1(t, m_i, m_j)$, 
in \Eq{cR} is given by \cite{Liu:2019rnx}
\e 
     \cR_1(t, m_i, m_j) = \int \hat{\cR}_1(t, m_i, m_j, m_l)\ \rd m_l,
\q 
where
\e 
\begin{split}
    &\hat{\cR}_1(t, m_i, m_j, m_l)
    =1.32 \times 10^6 \times \(\frac{t}{t_0}\)^{-\frac{34}{37}}\(\frac{\fpbh}{\mpbh}\)^\frac{53}{37} \\ 
 &\times m_l^{-\frac{21}{37}} (m_i m_j)^\frac{3}{37} (m_i + m_j)^\frac{36}{37} F(m_i) F(m_j) F(m_l).     
\end{split}
\q
The merger rate density of second-merger process, $\cR_2(t, m_i, m_j)$, 
in \Eq{cR} is given by \cite{Liu:2019rnx}
\e 
\begin{split}
     \cR_2(t, m_i, m_j) &= \hf \int \hat{\cR}_2(t, m_i-m_e, m_e, m_j, m_l)\ \rd m_l \rd m_e\\
    &+ \hf \int \hat{\cR}_2(t, m_j-m_e, m_e, m_i, m_l)\ \rd m_l \rd m_e,
\end{split}
\q 
where
\e 
\begin{split}
    &\hat{\cR}_2(t, m_i, m_j,m_k, m_l)
    =1.59 \times 10^4 \times \(\frac{t}{t_0}\)^{-\frac{31}{37}} \(\frac{\fpbh}{\mpbh}\)^\frac{69}{37}\\ 
 &\qquad\times m_k^{\frac{6}{37}} m_l^{-\frac{42}{37}} (m_i+ m_j)^\frac{6}{37} (m_i + m_j + m_k)^\frac{72}{37}\\
&\qquad\times F(m_i) F(m_j) F(m_k) F(m_l).     
\end{split}
\q
\section{\label{method}Inference on PBH mass distribution from GW data}
Given a general mass function of PBHs $P(m|\vth)$ which satisfy the normalization
condition of \Eq{norm}, the time (or redshift) dependent merger rate can be
obtained by \Eq{Rt}, namely
\e\label{Rt2}
R(t|\vth) = \int \mR(t, \vla|\vth)\ \rd \vla,
\q 
where $\vla \equiv \{m_1, m_2\}$, and $\vth$ are the parameters that 
characterize the mass function and will be inferred from GW data. 
For instance, $\vth=\{\alpha, M\}$ for the power-law PDF (see \Eq{power}) 
and $\vth=\{m_c, \s\}$ for the log-normal PDF (see \Eq{log}).
The local merger rate density distribution then reads \cite{Chen:2018rzo}
\e\label{Rt0} 
\mR(t_0, \vla|\vth) = R_0\, p(\vla|\vth),
\q 
where $R_0 \equiv R(t_0|\vth)$ is the local merger rate, and $p(m_1,m_2|\vth)$ 
is the population distribution of BBH mergers.
Note that \Eq{Rt0} guarantees $p(m_1,m_2|\vth)$ is normalized, namely
\e 
    \int p(\vla|\vth)\, \rd \vla = 1.
\q 

Given the GW data, $\vd = (d_1, \dots, d_N)$, which consist of $N$ BBH merger
events, we aim to extract the population parameters $\{\vth, R_0\}$ from $\vd$.
In order to do that, it is necessary to perform the hierarchical Bayesian inference
on the BBHs' mass distribution 
\cite{Abbott:2016nhf,Abbott:2016drs,TheLIGOScientific:2016pea,
    Wysocki:2018mpo,Fishbach:2018edt,Mandel:2018mve,Thrane:2018qnx}.
In this work, we will use the data of ten BBHs
\cite{TheLIGOScientific:2016pea,LIGOScientific:2018mvr} 
reported by \lvc\ O1 and O2 observations, and hence $N=10$. 
The posterior samples of these BBHs are public available from \footnote{\url{https://www.gw-openscience.org/}}. 
Because the standard priors on masses for each event in \lvc\ analysis are 
taken to be uniform \cite{TheLIGOScientific:2016pea,LIGOScientific:2018mvr},
the likelihood of an individual event $p(d_i|\vla)$ is
proportional to the posterior of that event $p(\vla|d_i)$.
The total likelihood for an inhomogeneous Poisson process can be evaluated as
\cite{Wysocki:2018mpo,Fishbach:2018edt,Mandel:2018mve,Thrane:2018qnx}
\e\label{likelihood}
p(\vd|\vth, R_0) \propto R_0^{N} e^{-R_0\, \beta(\vth)} \prod_i^N 
\int \rd\vla\ p(d_i|\vla)\ p(\vla|\vth),
\q 
where $\beta(\vth)$ is defined as
\e 
\beta(\vth) \equiv \int \rd\vla\ VT(\vla)\ p(\vla|\vth),
\q 
in which $VT(\vla)$ is the sensitive spacetime volume \cite{Abbott:2016nhf,Abbott:2016drs} of LIGO.
We adopt the semi-analytical approximation from 
\cite{Abbott:2016nhf,Abbott:2016drs} to estimate $VT$, 
where we use the ``IMRPhenomPv2" waveform to simulate the BBH templates
and neglect the effect of spins for BHs.
Furthermore, the threshold signal-to-noise ratio (SNR) of detection for a 
single-detector is set to $8$,
which corresponds to a network SNR threshold of around $12$.

Assuming the prior distributions $p(\vth, R_0)$ are uniform for $\vth$ parameters
and log-uniform for local merger rate $R_0$
\cite{Abbott:2016nhf,Abbott:2017vtc}, namely
\e 
p(\vth, R_0) \propto \frac{1}{R_0},
\q 
the posterior probability distribution $p(\vth, R_0|\vd)$ 
can be directly calculated by
\e\label{post} 
p(\vth, R_0|\vd) \propto p(\vd |\vth, R_0)\ p(\vth, R_0).
\q
The marginalized posterior $p(\vth|\vd)$ can then be readily obtained by
integrating over $R_0$ in \Eq{post}, namely
\e\label{post_vth} 
p(\vth|\vd) \propto \[\beta(\vth)\]^{-N} 
\prod_i^N \int \rd\vla\ p(d_i|\vla)\ p(\vla|\vth).
\q 
This marginalized posterior has been widely used in previous 
population inferences 
\cite{Abbott:2016nhf,Abbott:2017vtc,TheLIGOScientific:2016pea,
    Abbott:2016drs,Fishbach:2017zga,Chen:2018rzo}.
In the following section, we will utilize the posterior \eqref{post} to infer
the population parameters $\{\vth, R_0\}$ by considering two concrete mass
distributions, a power-law PDF and a log-normal PDF, respectively.

\section{\label{result}results}

\subsection{Power-law mass function}
We now consider a power-law mass function of PBHs as \cite{Carr:1975qj}
\e\label{power} 
    P(m) = \frac{\al-1}{M} \(\frac{m}{M}\)^{-\al},
\q
where $m>M$, and $\al>1$ is the power-law index.
Note that $\vth=\{\al, M\}$ and the free parameters are 
$\{\vth, R_0\} = \{\al, M, R_0\}$ in this case. 
Using \Eq{mpbh} and \Eq{Fm}, it is easily to get
\e
    \mpbh = M \frac{\al}{\al-1},
\q
\e 
    F(m) = \frac{\al}{m} \(\frac{m}{M}\)^{-\al}.
\q

\begin{figure}[htbp!]
	\centering
	\includegraphics[width = 0.48\textwidth]{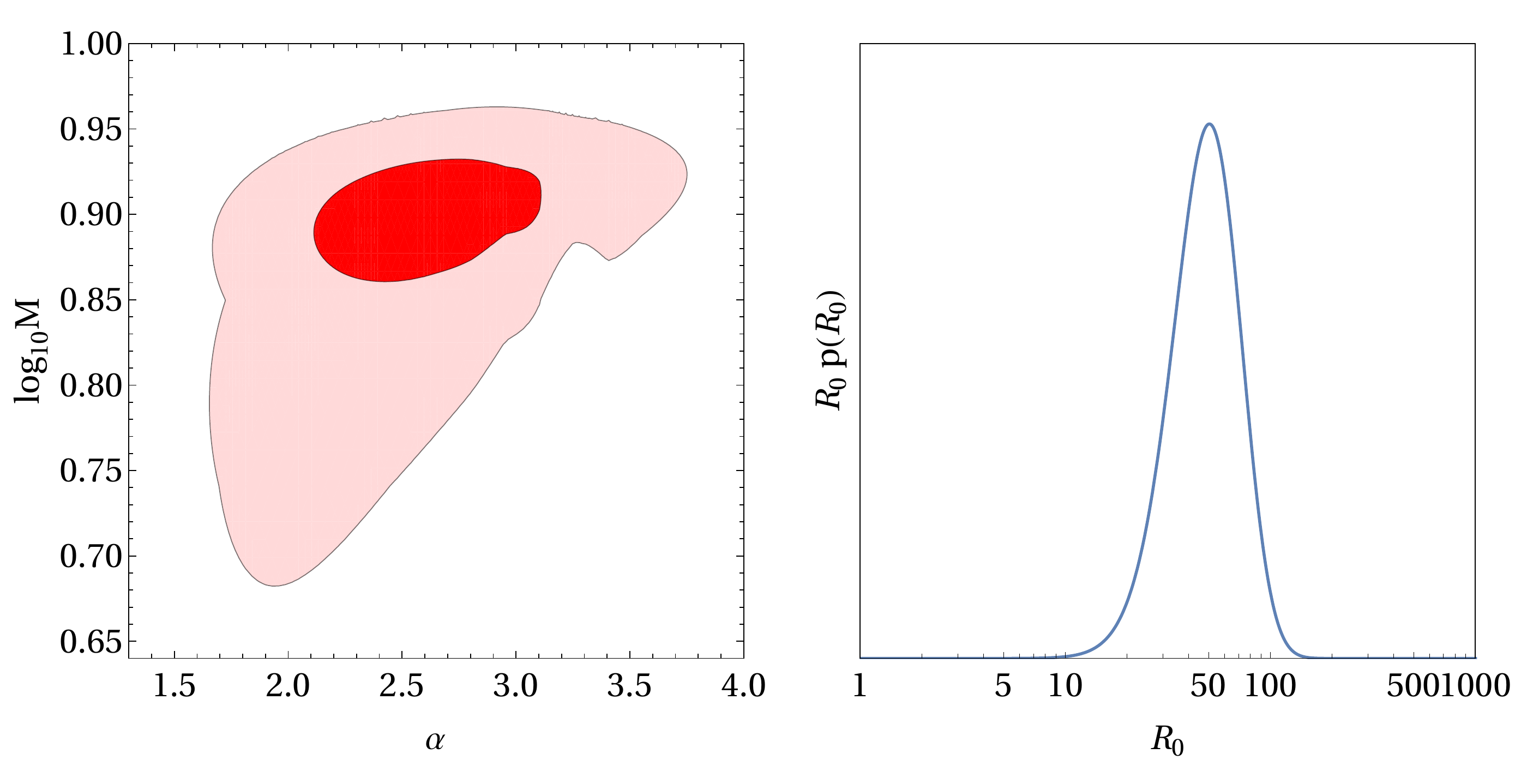}
	\caption{\label{posterior-power}
		The marginalized one- and two-dimensional posterior distributions for 
    parameters $\{\vth, R_0\} = \{\al, M, R_0\}$ in the power-law mass function 
    of PBHs, by using $10$ BBH events from \lvc\ O1 and O2 observing runs.
    The contours are at the $68\%$ and $95\%$ credible levels, respectively. 
	}
\end{figure}

Using data of $10$ BBHs observed by \lvc\ O1 and O2 observations and performing
the hierarchical Bayesian inference, we obtain $\al = 2.41^{+1.00}_{-0.87}$,
$M = 7.4^{+1.4}_{-3.3} \Msun$, and $R_0 = 48^{+37}_{-24} \gpcyr$.
It is then easy to infer the abundance of PBHs in CDM to be 
$\fpbh = 2.8^{+1.8}_{-1.2} \times 10^{-3}$ from the posterior distribution of local merger rate $R_0$.
The results of local merger rate and abundance of PBHs are consistent with
the previous estimations, confirming that the main components of CDM should
not be made of stellar mass PBHs.
The posteriors of parameters $\{\vth, R_0\} = \{\al, M, R_0\}$ are shown 
in \Fig{posterior-power}.

\begin{figure}[htbp!]
	\centering
	\includegraphics[width = 0.45\textwidth]{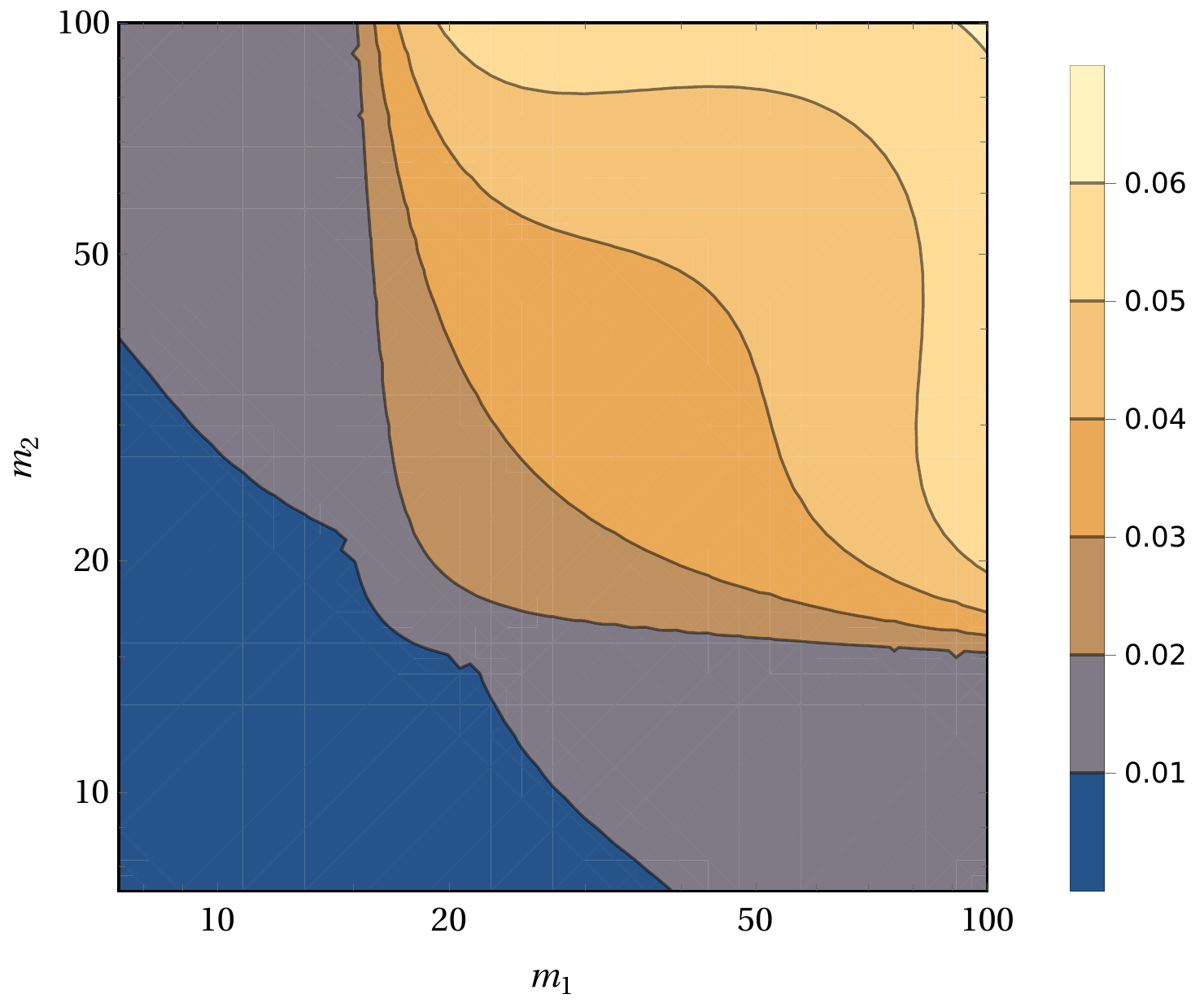}
	\caption{\label{ratio-power}
		The ratio of merger rate density from second-merger history
        to that from first-merger history,
        $\cR_2(t_0, m_1, m_2)/\cR_1(t_0, m_1, m_2)$.
	}
\end{figure}

\Fig{ratio-power} shows the ratio of merger rate density from second-merger history
to the one from first-merger history, namely 
$\cR_2(t_0, m_1, m_2)/\cR_1(t_0, m_1, m_2)$, by fixing $\{\vth, R_0\}$ to their best-fit values.
It is clearly that the correction of total merger rate density from 
second-merger history is less than $10\%$.
It is then readily to calculate the ratio of merger rate from 
second-merger history to the one from first-merger history, 
$R_2(t_0)/R_1(t_0) = 0.5\%$.
We therefore conclude that the merger history effect can be safely ignored
when estimating the merger rate (density) of PBHs.
\subsection{Log-normal mass function}
We now consider a log-normal mass function of PBHs as \cite{Dolgov:1992pu}
\e\label{log}
    P(m) = \frac{1}{\sqrt{2\pi} \s m} \exp \(-\frac{\ln^2\(m/m_c\)}{2\s^2}\),
\q
where $m_c$ presents the peak mass of $m P(m)$, and $\s$ denotes the width of 
the mass spectrum.
Note that $\vth=\{m_c, \s\}$ and the free parameters are 
$\{\vth, R_0\} = \{m_c, \s, R_0\}$ in this case. 
Using \Eq{mpbh} and \Eq{Fm}, it is easily to get
\e
    \mpbh = m_c \exp \(-\frac{\s^2}{2}\),
\q
\e 
    F(m) = \frac{m_c}{\sqrt{2\pi} \s m^2} \exp \(-\frac{\s^2}{2}-\frac{\ln^2\(m/m_c\)}{2\s^2}\).
\q

\begin{figure}[htbp!]
	\centering
	\includegraphics[width = 0.48\textwidth]{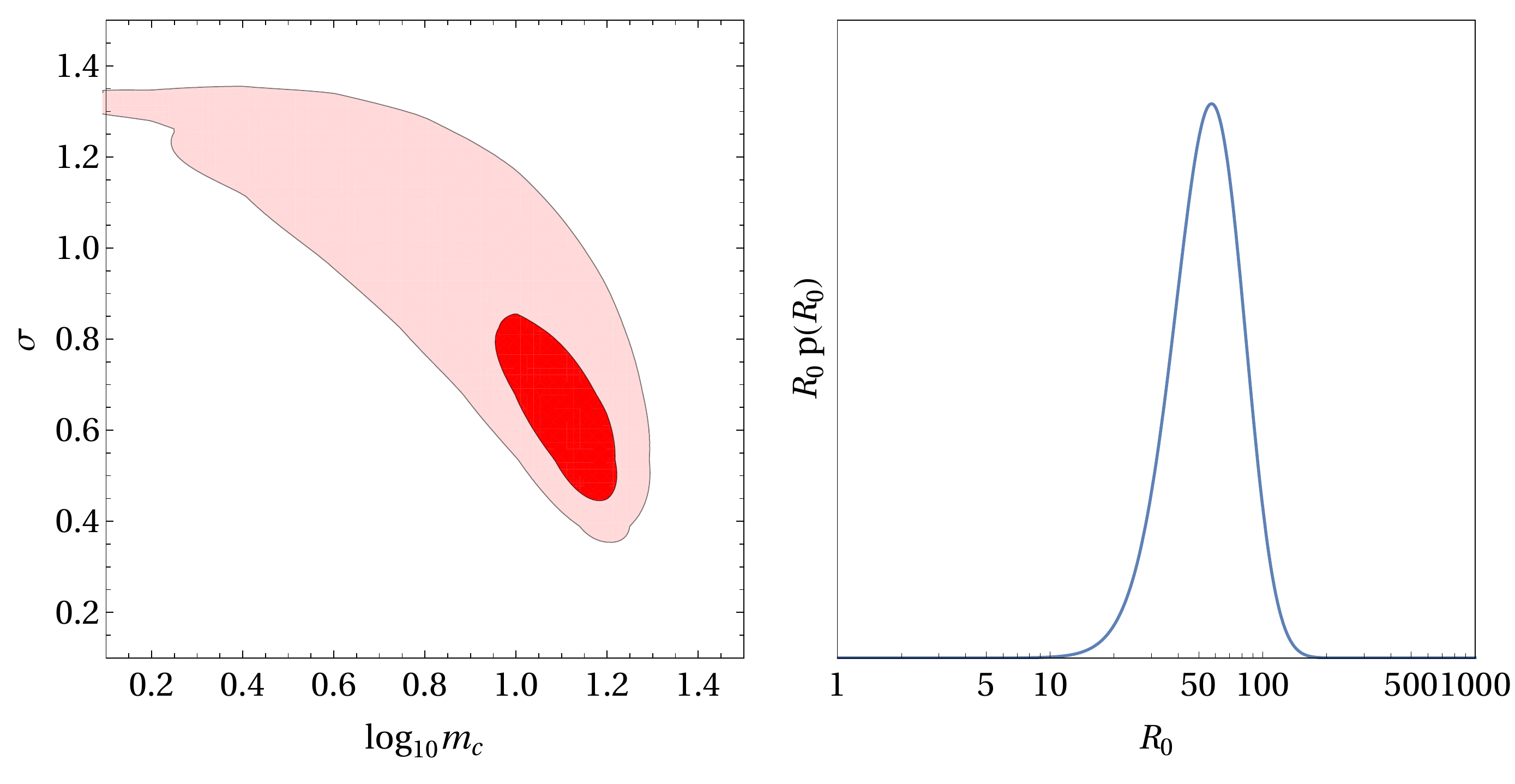}
	\caption{\label{posterior-log}
		The marginalized one- and two-dimensional posterior distributions for 
    parameters $\{\vth, R_0\} = \{m_c, \s, R_0\}$ in the power-law mass function 
    of PBHs, by using $10$ BBH events from \lvc\ O1 and O2 observing runs.
    The contours are at the $68\%$ and $95\%$ credible levels, respectively. 
	}
\end{figure}

Using data of $10$ BBHs observed by \lvc\ O1 and O2 observations and performing
the hierarchical Bayesian inference, we obtain $m_c =8.9^{+7.8}_{-7.3} \Msun$,
$\s = 0.91^{+0.50}_{-0.42}$, and $R_0 = 55^{+42}_{-27} \gpcyr$.
It is then easy to infer the abundance of PBHs in CDM to be 
$\fpbh = 2.6^{+6.8}_{-1.4} \times 10^{-3}$ from the posterior distribution of 
local merger rate $R_0$.
The results of local merger rate and abundance of PBHs are consistent with
the previous estimations, confirming that the main components of CDM should
not be made of stellar mass PBHs.
The posteriors of parameters $\{\vth, R_0\} = \{m_c, \s, R_0\}$ are shown 
in \Fig{posterior-log}.

\begin{figure}[htbp!]
	\centering
	\includegraphics[width = 0.45\textwidth]{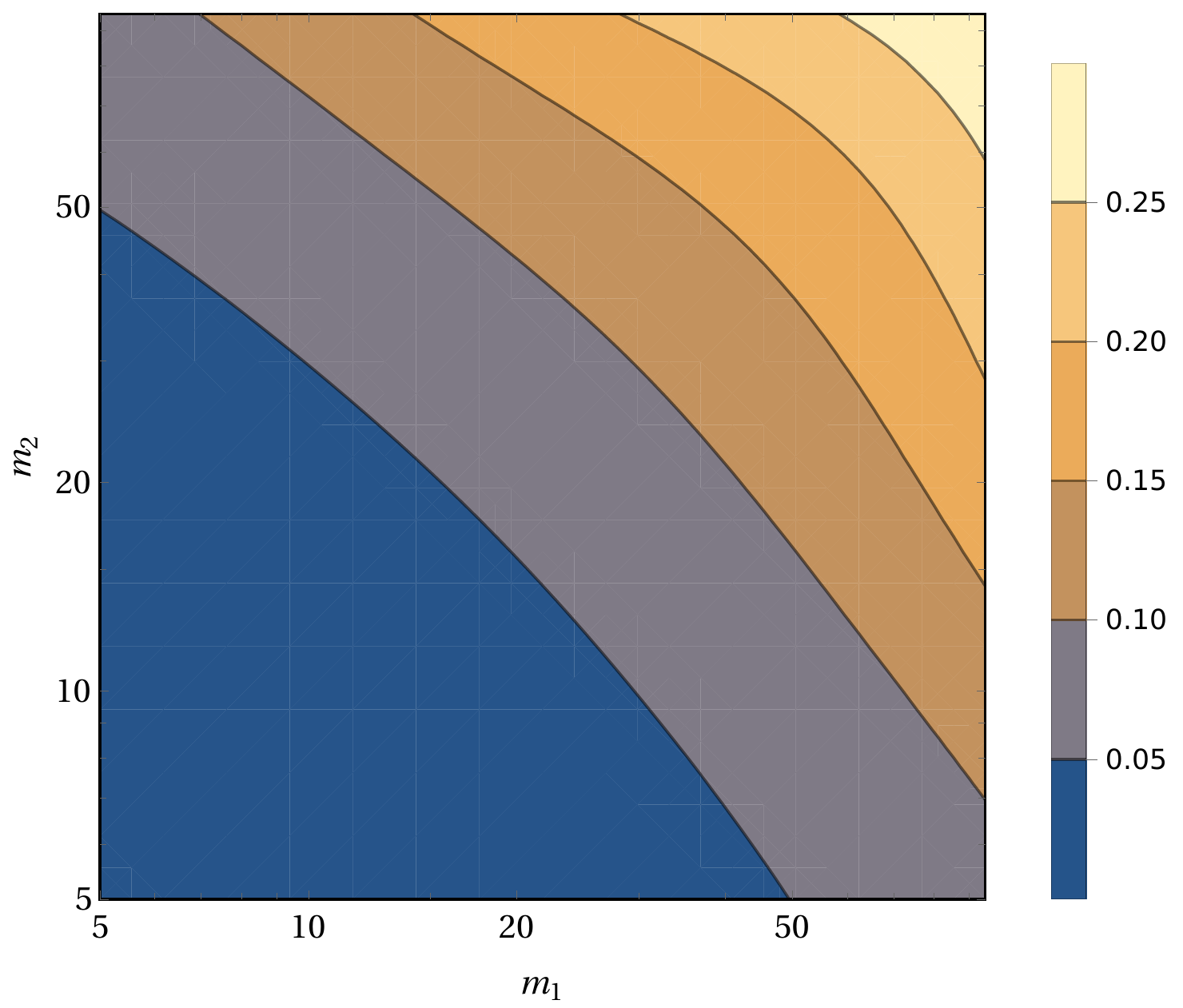}
	\caption{\label{ratio-log}
		The ratio of merger rate density from second-merger history
        to that from first-merger history,
        $\cR_2(t_0, m_1, m_2)/\cR_1(t_0, m_1, m_2)$.
	}
\end{figure}

\Fig{ratio-log} shows the ratio of merger rate density from second-merger history
to the one from first-merger history, namely 
$\cR_2(t_0, m_1, m_2)/\cR_1(t_0, m_1, m_2)$, by fixing $\{\vth, R_0\}$ to their best-fit values.
The correction to total merger rate density from 
second-merger history is larger as component masses are heavier.
However, the ratio of merger rate from 
second-merger history to the one from first-merger history is negligible, namely
$R_2(t_0)/R_1(t_0) = 3.0\%$.
This is because the major contribution to the merger rate are from the binaries 
with masses less than $50\Msun$.
Therefore the merger history effect can be safely ignored
when estimating the merger rate of PBHs.

\section{\label{conclusion}Conclusion}

In this paper, we use the public available GW data of $10$ BBH events from 
\lvc\ O1 and O2 observing runs to constrain the merger rate distribution 
of PBHs by accounting the effect of merger history.
Considering two concrete mass functions of PBHs, a power-law PDF and 
a log-normal one, we demonstrate that the contribution of merger rate (density)
from second-merger history to total merger rate (density) is 
subdominant, and hence the second-merger history effect can be safely ignored.
As third-merger (and later merger) history will make even less contribution to 
the total merger rate (density), we conclude that the effect of merger history
is subdominant and can be neglected when evaluating the merger rate of PBH
binaries.

Furthermore, the results of local merger rate and abundance of PBHs inferred from the updated
analysis are consistent with the previous estimations, 
confirming that the main components of CDM should not be made of stellar 
mass PBHs. 

\begin{acknowledgments}
We would like to thank Zu-Cheng Chen and Lang Liu for useful conversations.
This research has made use of data, software and/or web tools obtained 
from the Gravitational Wave Open Science Center \cite{Vallisneri:2014vxa}
(\url{https://www.gw-openscience.org}), a service of LIGO Laboratory, 
the LIGO Scientific Collaboration and the Virgo Collaboration. 
LIGO is funded by the U.S. National Science Foundation. 
Virgo is funded by the French Centre National de Recherche Scientifique (CNRS),
the Italian Istituto Nazionale della Fisica Nucleare (INFN) and the Dutch Nikhef,
with contributions by Polish and Hungarian institutes. 
\end{acknowledgments}	
	
\bibliography{./ref}

\end{document}